\documentclass[10pt,twocolumn,letterpaper]{article}
\usepackage{ijcb}
\usepackage{times}
\usepackage{epsfig}
\usepackage{graphicx}
\usepackage{amsmath}
\usepackage{amssymb}
\usepackage{hyperref}
\usepackage{subfigure}
\usepackage{amssymb}
\usepackage[ruled,vlined]{algorithm2e}
\usepackage{multirow}
\usepackage{authblk}
\usepackage[norule, perpage, symbol]{footmisc}
\usepackage[numbers,sort&compress]{natbib}

\emergencystretch=\maxdimen
\ijcbfinalcopy 

\ifijcbfinal\fi

\makeatletter
\def\ps@IEEEtitlepagestyle{
\def\@oddfoot{\mycopyrightnotice}
\def\@evenfoot{}
}
\def\mycopyrightnotice{
{\footnotesize 
2021 IEEE International Joint Conference on Biometrics (IJCB) \\ \hfill 978-1-6654-3780-6/21/\$31.00 \copyright 2021 IEEE\hfill}
}
\makeatother

\begin{document}
\title{Defending Touch-based Continuous Authentication Systems from Active Adversaries Using Generative Adversarial Networks}
\author[1]{Mohit Agrawal*}
\author[2]{Pragyan Mehrotra*}
\author[3]{Rajesh Kumar}
\author[2]{Rajiv Ratn Shah}
\affil[1]{Teradata R\&D Labs, India. \textsuperscript{2}IIIT Delhi, India. \textsuperscript{3}Haverford College, USA.}
\affil[ ]{\text{\href{mailto:mohit.agrawal@teradata.com}{mohit.agrawal@teradata.com}, \href{mailto:pragyan18168@iiitd.ac.in}{pragyan18168@iiitd.ac.in}, \href{mailto:rkumar@haverford.edu}{rkumar@haverford.edu}, \href{mailto:rajivratn@iiitd.ac.in}{rajivratn@iiitd.ac.in}}}

\maketitle
\phantomsection
\thispagestyle{empty}
\footnotetext{*Authors contributed equally.}

\begin{abstract}
Previous studies have demonstrated that commonly studied (vanilla) touch-based continuous authentication systems (V-TCAS) are susceptible to population attack. This paper proposes a novel Generative Adversarial Network assisted TCAS (G-TCAS) framework, which showed more resilience to the population attack. G-TCAS framework was tested on a dataset of $117$ users who interacted with a smartphone and tablet pair. On average, the increase in the false accept rates (FARs) for V-TCAS was much higher ($22\%$) than G-TCAS ($13\%$) for the smartphone. Likewise, the increase in the FARs for V-TCAS was $25\%$ compared to G-TCAS ($6\%$) for the tablet. 
\end{abstract}
\let\thefootnote\relax\footnotetext{\mycopyrightnotice}
\section{Introduction}
Individuals' identity can be authenticated via what one: can memorize (\eg, PIN, passwords), can carry (\eg., magnetic cards, keyfobs), has (\eg, face, fingerprint, iris), and does (\eg, walking, talking, swiping). These means for authentication offer advantages over one another primarily in usability, privacy, and security. For instance, PINs and passwords need to be memorized, are time-consuming, or can be stolen \cite{ShuklaCCS14, ShuklaTIFS2020}. Face or fingerprint need not be memorized and offer faster authentication. However, some users might find them intrusive. Similarly, swiping or typing patterns could offer continuous authentication of identity, while PIN, password, and fingerprint offer only entry-point authentication. Touch-gestures are among the most widely studied authentication means for continuous authentication \cite{frank2012touchalytics,TouchFirstAuth,Li2013UnobservableRF,serwadda2013verifiers,ModalSwipeContinuous, KeystrokeTimingMovement2016, patel2016continuous, primo2014context, zhang}. One of the primary reasons behind this is that touch gestures meet most of the criteria (universality, distinctiveness, permanence, collectability, performance, acceptability, and circumvention) that are defined to be viable biometric \cite{LatestSurveyTouchUsabilitySecurity2020, BiometricCriteria}. 

Hertenstein and Keltner \cite{TouchCommunicateEmotions} stated that \textit{touch} is the most developed sensory modality since birth and contributes to people's cognitive and socio-emotional development. They suggest that people can decode anger, fear, disgust, love, gratitude, and sympathy via touch. One can even estimate the length of thumbs from touch gestures produced on touch-enabled smartphones \cite{SwipeToHandGeometry}. The authors cited anthropometrics to argue that many individual body segments' lengths follow a unique proportional relationship. The pervasiveness of touch-enabled devices allows us to capture touch from different dimensions, including the touch locations and area and pressure at each of those locations. Numerous studies \cite{frank2012touchalytics, TouchFirstAuth, Li2013UnobservableRF, ModalSwipeContinuous, KeystrokeTimingMovement2016} have demonstrated that commonly studied (vanilla) touch-based continuous authentication systems (V-TCAS) achieve practically low error rates. However, most of these studies have evaluated V-TCAS's performance under a zero-effort adversarial environment despite some studies suggesting that V-TCAS are vulnerable to data injection and imitation attacks \cite{RoboticRobbery, MimicryAttackOnSwipes, RandomAttackMutibiometric}. 

We believe that it is essential to take a proactive approach and test TCAS under adversarial environments in the literature. This paper makes an effort in this direction. The main contributions are summarized as follows: 
\begin{itemize}
    \item We implement and test the vanilla TCAS (V-TCAS) under traditional zero-effort and population-based adversarial scenarios. In line with previous studies, we observed that V-TCAS's false acceptance increases significantly under the population-based adversarial scenarios.
    
    \item Next, we propose a novel Generative Adversarial Networks assisted TCAS framework (G-TCAS) and test the same under zero-effort and population attack adversarial scenarios and found it to be more resilient than V-TCAS.
    
    \item We benchmarked four widely studied classifiers (each with a diverse learning paradigm) on a dataset of $117$ users who provided their samples in a multi-device environment. The superiority of G-TCAS was evident over V-TCAS across the classifiers and devices\textsuperscript{1}\footnote{\textsuperscript{1} \href{https://github.com/midas-research/IJCB2021GANTouch}{https://github.com/midas-research/IJCB2021GANTouch}}. 
    
\end{itemize}

The rest of the paper is organized as follows. Section \ref{RelatedWorks} discusses the closely related works. Section \ref{DesignOfExperiments} presents the design of experiments. Section \ref{ResultsAndDiscussion} presents and discusses the results, respectively. Finally, we conclude the paper and provide future research directions in Section \ref{secConclusionAndFutureWork}.

\section{Related work}
\label{RelatedWorks}
The closely related works are structured as follows. The first part discusses the studies that focus on TCAS. While the second part systematically described various possible adversarial environments.
\subsection{Continuous Authentication via Touch Gestures}
Some of the earliest studies that explored touch gesture for authentication include \cite{frank2012touchalytics,TouchFirstAuth,Li2013UnobservableRF,serwadda2013verifiers, ModalSwipeContinuous}. These studies focused primarily on collecting data and demonstrating that touch gestures are unique to an individual under the non-existent active adversary's assumption. Later studies \cite{KeystrokeTimingMovement2016,serwadda2013verifiers,2018Benchmark} divided the swipes into multiple types such as left, right, up, and down; built a separate model for each of these types to authenticate users. Kumar \etal. \cite{KeystrokeTimingMovement2016} studied touch gestures and corresponding movements captured by an accelerometer. They argued that it would be a little more difficult for adversaries to reproduce both the touch gestures and the underlying movements at any given point in time than just spoofing the swipes.

The majority of the previous studies focused on analyzing touch gestures collected from smartphones \cite{frank2012touchalytics, sitova2015hmog, TouchFirstAuth, RamaChellappa, chellappa2019continuous, serwadda2013verifiers, Li2013UnobservableRF, KeystrokeTimingMovement2016}. These studies collect data from human participants while they answered a set of questions, browsed websites, or scrolled images. Later, they extracted a list of features from raw touch events, usually consisted of touch coordinates, time of touch events, pressure, and area along with those coordinates. The extracted features were used to train the authentication models. The training part consisted of training of one \cite{OneClassRajesh2018,saravanan2014latentgesture} or two class classifiers \cite{frank2012touchalytics,serwadda2013verifiers,2018Benchmark}. While the testing part focused on passing genuine and non-genuine samples through the model and computing the genuine fail (false reject rates) and impostor pass rates (false accept rates). The majority of the studies have used these two metrics to report the performance of TCAS. Some studies have also reported Equal Error Rate \cite{frank2012touchalytics} and Half Total Error Rates (an average of false accept and false reject rates) \cite{sitova2015hmog}. Researchers have recommended the use of Half Total Error Rates to report the testing performance as it is not possible to change the threshold during testing \cite{bengio2002confidence}. 

The majority of the studies that use touch gestures collected from smartphones have reported error rates under $10\%$ percent, which is an accepted region, especially for continuous authentication. A continuous authentication system with $10\%$ or lower error rates has several applications in the civilian domain. The problem, however, is that the studies have assumed the non-existence of active adversaries or people with malicious intent. Since the data generated or stored on smart devices are invaluable, there is a high likelihood that active adversaries would attack these systems. Therefore, this study focuses on the evaluation of TCAS under an active adversarial environment and proposing countermeasures. 

Apart from smartphones, one of the heavily used touch-enabled devices is the tablet. Thus, a few studies have explored touch gestures collected via tablet for continuous authentication \cite{saravanan2014latentgesture, trojahn2013toward}. Saravanan \etal. \cite{saravanan2014latentgesture} trained LibSVM, a one-class classifier, and obtained an average accuracy of $97.9\%$ and $96.79\%$ on the smartphone and tablet, respectively, on a small dataset of $20$ users. While Trojahn \etal. \cite{trojahn2013toward} analyzed the slide-touch-type and reported as low as $2\%$ error rates on a dataset of $18$ users. It is worthwhile to note that not many studies have explored smartphones and tablets together, especially keeping the users common across the devices. 

\subsection{Adversarial Frameworks for TCAS} The idea of adversarial frameworks or attack models is still developing with a common goal, \ie, to fool the authentication system. Based on the current literature, the adversarial frameworks can be divided into two categories, namely, \textit{data-injection-based} \cite{RandomAttackMutibiometric} and \textit{imitation-based} \cite{MimicryAttackOnSwipes,RoboticRobbery}. 

The \textit{data-injection-based} attacks assume that it would be possible to bypass the link between the sensor and the authentication Application Programming Interface (API). Thus, the attackers inject snooped, spoofed, random or population-derived samples into the authentication API such that the API will grant (or keep granting) the access \cite{RandomAttackMutibiometric}. The data-injection-based attacks could be grouped into random \cite{RandomAttackMutibiometric}, user-tailored (from keystrokes \cite{Snoop-forge-reply-keystroke}), and population-based attacks (from keystrokes \cite{SerwaddaPopulation}). The random attacks include the injection of samples collected from random individuals or random samples generated statistically. The user-tailored attacks include the injection of samples stolen or estimated for a targeted user (popular as snoop-forge-replay attacks). In comparison, population-based attacks inject the average samples (similar to average face) derived from a large public dataset. Needless to say, that population attack eliminates the requirements of stealing the sample from the genuine user.

On the other hand, the \textit{imitation-based} attacks use live samples collected from humans, robots, or human plus robots. Based on the amount and method of training the individuals or robot, the imitation-based attacks fall into zero-effort or high-effort categories. The zero-effort refers to the scenarios in which attackers use random users who produce the touch gestures without any explicit training or attempt to copy \cite{frank2012touchalytics, sitova2015hmog, TouchFirstAuth, RamaChellappa, chellappa2019continuous, serwadda2013verifiers, Li2013UnobservableRF, KeystrokeTimingMovement2016}. While under the high-effort, the attackers either train an individual or robot to mimic targeted individual or average gestures derived from publicly available databases \cite{MimicryAttackOnSwipes, RoboticRobbery}. The imitation-based attacks are realistic but tedious and require a much higher level of effort than the data-injection attacks because they eliminate the assumption of bypassing the sensor to API link. In this study, we focus on the \textit{data-injection-based} attack frameworks and leave the \textit{imitation-based} attacks for the future, as they require much more resources to realize (\eg designing a robot).

It is important to note that a biometric system like TCAS could be attacked from many different sources as described in \cite{AttackSources}. The attack methods other than the ones described earlier would require write permission to the authentication API. The authentication API's implementation is supposed to be secure, leaving other sources of attack almost impractical. Therefore, researchers have focused mostly on data-injection- and imitation-based attacks. The success of these attacks depends on the resilience of the pattern matching component in the authentication API. Researchers have noted that raw data level distance-based matching is more resilient to the attacks; however, they exhibit very high error rates, in general, \cite{RandomAttackMutibiometric, DistanceBasedMatchersAreImmuneToRandomAttacks,rajesh2020,rajesh2015}. Machine learning-based matchers, on the other hand, achieve much lower error rates, therefore, heavily studied for implementing TCAS \cite{2018Benchmark, LatestSurveyTouchUsabilitySecurity2020, serwadda2013verifiers} than the distance-based matchers. Therefore, we chose to experiment with machine learning-based implementations of TCAS. To summarize, we evaluate the most widely studied implementations of TCAS under the most common adversarial scenarios. 

\section{Design of Experiments}
This section details the data collection, feature engineering, feature analysis, and implementations of V-TCAS and G-TCAS. 

\label{DesignOfExperiments}
\subsection{The Datasets}
We used a public dataset named Syracuse University and Assured Information Security-Behavioral Biometrics Multi-Device, and Multi-Activity Data (SU-AIS BB-MAS) \cite{belman2019insights}. SU-AIS BB-MAS consists of multiple modalities like keystroke, touch, and gait; however, we consider only the touch part, therefore, refer to the dataset as BBMAS-Touch throughout this paper. The reasons to choose BBMAS-Touch include the number of users, the number of samples per user, and common users across multiple devices. Additionally, we used another publicly available dataset \cite{serwadda2013verifiers} to create a more realistic population-based attack environment and refer to the same as Serwadda-Touch. 

Both BBMAS-Touch and Serwadda-Touch datasets consisted of raw touch information (coordinates, pressure, and area at every touchpoint) collected while participants answered a series of questions in two different sessions. The questions were designed in such a way that they generated both horizontal and vertical swipes. More details in the data and data collection experiment are available in \cite{serwadda2013verifiers} and \cite{belman2019insights, Vishaal2020}. 

\begin{table}[htp]
\footnotesize
\caption{List of features gleaned from individual swipes \cite{frank2012touchalytics}.}
\begin{tabular}{|c|c|}
\hline
\multicolumn{1}{|c|}{\textbf{Feature}}                                                        & \multicolumn{1}{c|}{\textbf{Description}}                                                   \\ \hline
$swipe\_duration$                                                                               & $t_{end}$ - $t_{start}$                                                                               \\ \hline

\begin{tabular}[c]{@{}l@{}}$start\_x$, $start\_y$, \\ $end\_x$, $end\_y$ \end{tabular}                 & $x_0$, $x_0$, $x_{n-1}, $y$_{n-1}$

\\ \hline
$dp$                                                                                             & Displacement of swipe                                                         \\ \hline
$l$                                                                                             & Length of the swipe                                                                         \\ \hline
$velocity$                                                                                      & $dp$/($t_{end} - t_{start}$)                                                              \\ \hline
$initial\_v$                                                                                    & Initial velocity (first 5\% of the points)                                                  \\ \hline
$final\_v$                                                                                      & Final Velocity (final 5\% of the points)                                                    \\ \hline
$mean\_v$                                                                                       & Pairwise average velocity (magnitude)                                                       \\ \hline
$direction$                                                                                     & Slope of line joining start and end points                                                  \\ \hline
$area$                                                                                          & average area of the fingertip over the swipe                                                \\ \hline
$acceleration$                                                                                  & Acceleration between start and end points                                                   \\ \hline
$mean\_a$                                                                                       & pairwise average acceleration (magnitude)                                                   \\ \hline
$initial\_a$                                                                                    & Initial acceleration (first 5\% points)                                                     \\ \hline
$final\_a$                                                                                      & Final acceleration (final 5\% points)                                                       \\ \hline
$aP_{25}$, $aP_{50}$, $aP_{75}$                                                                              & Acceleration Percentile ($aP_m$) at $m$\% swipe                                                  \\ \hline
$vP_{25}$, $vP_{50}$, $vP_{75}$                                                                              & Velocity Percentile ($vP_m$) at $m$\% swipe                                                      \\ \hline
$speed$                                                                                         & $l$/($t_{end} - t_{start}$)                                                                           \\ \hline
$initial\_s$, $final\_s$                                                                          & Initial Speed, Final Speed                                                                  \\ \hline
$sP_{25}$, $sP_{50}$, $sP_{75}$                                                                              & Speed Percentile ($sP_m$) at $m$\% swipe                                                         \\ \hline
\begin{tabular}[c]{@{}l@{}}$mean\_v_x$, $mean\_v_y$, \\ $mean\_a_x$, $mean\_a_y$, \\ $mean\_d$\end{tabular} & Average of $v_x$, $v_y$, $a_x$, $a_y$, $dp$                                                            \\ \hline
$max\_d$                                                                                        & Maximum of deviations                                                                       \\ \hline
$v_xP_{25}$, $v_xP_{50}$, $v_xP_{75}$                                                                           & \begin{tabular}[c]{@{}l@{}}Mean Velocity Percentile ($v_xP_m$) \\ at $m$\% swipe\end{tabular}     \\ \hline
$v_yP_{25}$, $v_yP_{50}$, $v_yP_{75}$                                                                           & \begin{tabular}[c]{@{}l@{}}Mean Velocity Percentile ($v_yP_m$) \\ at $m$\% swipe\end{tabular}     \\ \hline
$a_xP_{25}$, $a_xP_{50}$, $a_xP_{75}$                                                                           & \begin{tabular}[c]{@{}l@{}}Mean Acceleration Percentile ($a_xP_m$) \\ at $m$\% swipe\end{tabular} \\ \hline
$a_yP_{25}$, $a_yP_{50}$, $a_yP_{75}$                                                                           & \begin{tabular}[c]{@{}l@{}}Mean Acceleration Percentile ($a_yP_m$) \\ at $m$\% swipe\end{tabular} \\ \hline
\end{tabular}
\label{touchfeatures}
\end{table}

\subsection{Feature Engineering and Analysis} As mentioned earlier, comparison of raw data does not result in acceptable error rates. Therefore, almost all previous studies have relied on the extraction of features from the raw touch events. The derived features have been effectively used to train classifiers that successfully distinguished between genuine and non-genuine (impostor) users. Following previous studies, we extracted features from individual swipes (a swipe here is a series of touch events after finger down, during the slide, and till finger up events). The outlying swipes (with less than five touch points) were removed from the database. A swipe with five or fewer points can be considered a tap. We decided to exclude them from our study because those events were not found unique enough among the users during exploratory swipe analysis. The numbers of total swipes extracted from the phone and tablet were 22625 and 18527, respectively. The outlier removal process resulted in the exclusion of 2339 (10.33\%) and 4713 (25.43\%) swipes. We chose to include swipes with longer latency in contrast to \cite{sitova2015hmog} because the data collection experiments consisted of questions with varying cognitive load, resulting in longer duration swipes. In other words, longer swipes could be someone's unique trait. We extracted the $30$ features used in Frank \etal. \cite{frank2012touchalytics} from each swipe, and added $17$ more features which were variations of the idea around the thirty features. As a result, we used $47$ features in total from each swipe as listed in Table \ref{touchfeatures}.  

A swipe $S$ is a culmination of $n$ touch events. It can be represented as a tuple given in Equation \ref{e0}. 

\begin{equation}\label{e0}
S = (x, y, t, a, b)_{\texttt{i=1 to n}}
\end{equation}
where $x, y, t, a,$ and $b$ represent, x-coordinate, y-coordinate, time, and major-axis and minor-axis of the fingertip of each touch event, respectively. 

Pairwise velocities ($v_x$ and $v_y$) for x and y axes, $\forall i \in [1, n)$ were computed as described in Equation \ref{e1}.
\begin{equation}\label{e1}
(v_x)_i =  \frac{x_i - x_{i-1} }{t_i - t_{i-1}}, (v_y)_i =  \frac{y_i - y_{i-1} }{t_i - t_{i-1}}   
\end{equation}
Similarly, pairwise accelerations $(a_x$ and $a_y)$, length of the swipe ($l$), mean area of the fingertip ($A$), can be calculated as shown in Equations \ref{e2}, \ref{e3}, \ref{e4}, respectively.

\begin{equation}\label{e2}
(a_x)_i = { \frac{(v_x)_{i} - (v_x)_{i-1} }{t_i - t_{i-1}}}, (a_y)_i = { \frac{(v_y)_{i} - (v_y)_{i-1} }{t_i - t_{i-1}}}  
\end{equation}\

\begin{equation}\label{e3}
l = \sum_{i = 1}^{ n - 1} \sqrt{(x_{i-1} - x_i)^2 + (y_{i-1} - y_i)^2} 
\end{equation}

\begin{equation}\label{e4}
A = \frac{1}{n}\sum_{i=1}^{n}\pi \times a_i \times b_i
\end{equation}

Meanwhile, the deviations of each point, defined as the distance from the line joining the starting and ending point of the swipe ($d_i$), can be calculated with Equation \ref{e5}.

\begin{equation}\label{e5}
d_i = \frac{|y_i - m \times x_i - c|}{\sqrt{1 + m^2}}
\end{equation}
Assuming the equation of the line to be $y = m \times x + c$ 

The velocity, acceleration, and speed features were computed at first, second, and third quantiles of the swipe's touch events. These features capture the live nature of the swipes. Similarly, velocities and accelerations at the beginning and end of the swipe help us distinguish between the users who start/end their swipe aggressively or gently.

\subsection{Authentication Framework} 
\begin{figure*}[htp]
    \centering
    \includegraphics[height=3.45in, width=6.7in]{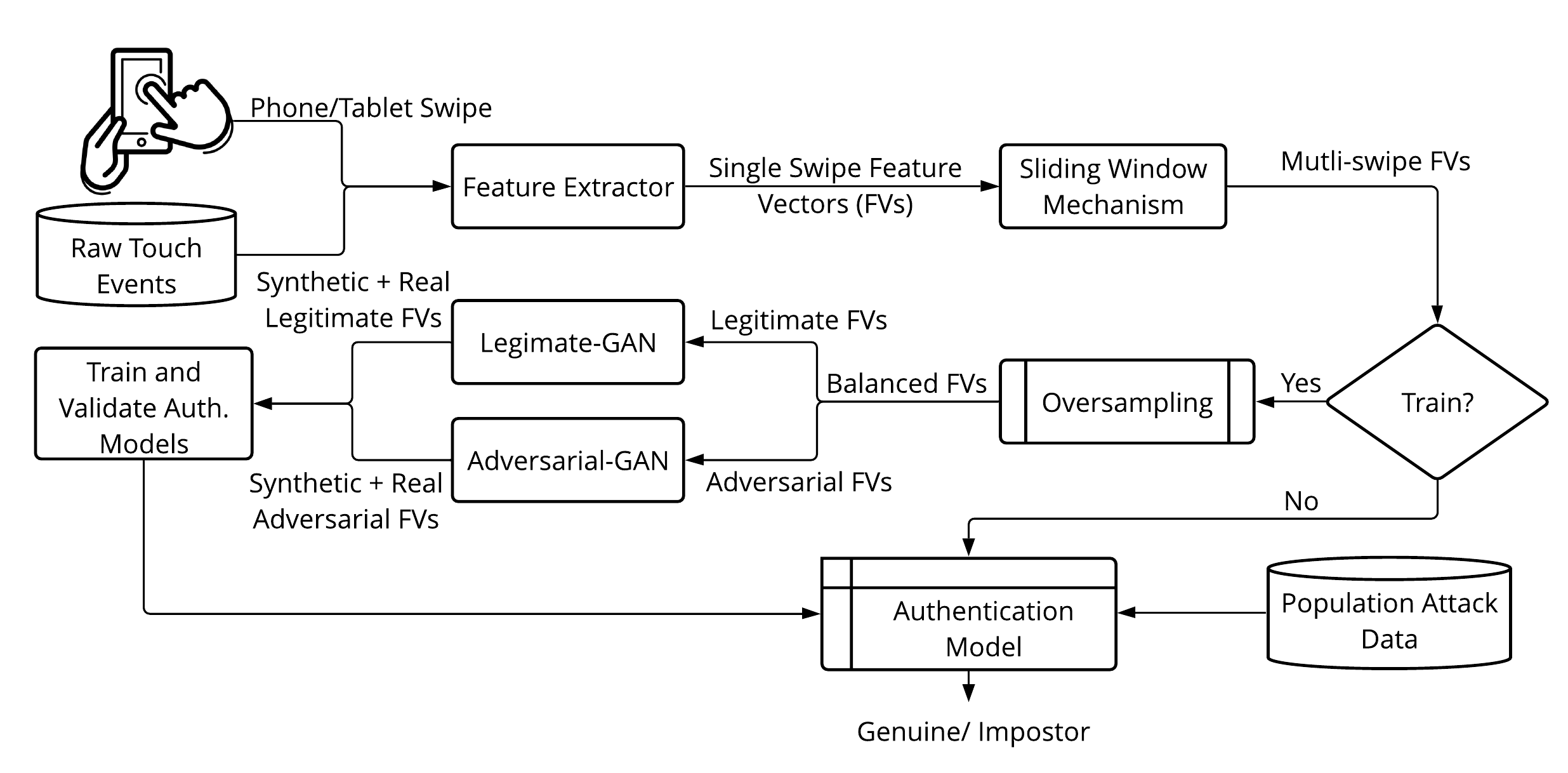}
    \caption{\small The system architecture of TCAS that uses Legitimate and Adversarial Generative Adversarial Networks. } 
    \label{adversarialtouch}
\end{figure*}

The authentication framework depicted in Figure \ref{adversarialtouch} consisted of the following components.  
\vspace{0.1in}

\textbf{The training and testing datasets.} The dataset was split into train and test sets with $60\%$ and $40\%$ of swipes for every user, respectively. The train set was used for parameter tuning with $k$-fold cross-validation ($k=5$). The test set was kept separate (unseen) and was used only during the testing process to mimic a real-world setup.
\vspace{0.1in}

\textbf{Continuous authentication.} One of the simplest ways to achieve continuous authentication via touch gestures is to use sliding windows of touch gestures and provide authentication decisions for each window. The design goal generally is to minimize both the window size and the sliding interval. The former decides the time taken in the first authentication decision, while the latter dictates the time taken by the subsequent authentication decision. We took $p$ consecutive swipes in one window with the removal and addition of $q$ swipes for the next window. A range of values for $p$ and $q$ were inspected. $p=5$ and $q=1$ achieved the best training error rates, in turn, adapted for training the final models and testing.  
\vspace{0.1in}

\textbf{Choice of classification algorithms.} We applied three criteria to choose the classifiers for our study. First, the classifiers must have been used successfully (achieved less than $10\%$ error rates) in the past studies. Second, the classifier should not be data-hungry for training, primarily because most public datasets have a small number of touch gestures per user. The third criterion was the diversity of learning paradigms. The process resulted in the selection of Support Vector Machine (SVM), Random Forest (R.F.), and Multilayer Perceptron (MLP) \cite{frank2012touchalytics, 2018Benchmark, serwadda2013verifiers, RoboticRobbery}. We added Extreme Gradient Boosting (XGB), which has not been studied much in the touch biometric domain but has performed well in other domains. 
\vspace{0.1in}

\textbf{The class imbalance.} The two-class classifiers require samples from both classes to be trained. To meet this requirement, traditionally, researchers have used the genuine users' feature vectors as genuine samples and feature vectors of users other than the genuine users as impostor samples. However, this strategy results in a heavy class imbalance as the number of impostor samples is always higher. Past studies have used under-sampling, over-sampling, or both to address the problem. In this paper, first, we take only a limited number (four from each possible impostor) of samples from the rest of the users, so the number of impostor samples is $4 \times (n-1)$, where $n$ is the total number of users in the dataset. The process of not choosing all the samples from all possible impostors can be termed as under-sampling. At the same time, we employed an Adaptive synthetic sampling approach for imbalanced learning (ADASYN) \cite{he2008adasyn} to over-sample the genuine feature vectors. As a result, every training and testing activity in our experiment used an equal number of genuine and impostor samples. 
\vspace{0.1in}

\begin{algorithm} 
\small
\SetAlgoLined
\textbf{Input:} ($X, n$) $X:$ combined feature matrix of all users in the attack dataset, $n:$ is the number of population attack vectors to be generated.\\
\textbf{Output:} $X':$ set of $n$ population attack vectors.\\

 $\mu$ $\gets$ [$mean(X_i)$ for $X_i$ $\in$ X.cols]\\
 $\sigma \gets$ [$std(X_i)$ for $X_i \in$ X.cols]\\
 $X' \gets$ []\\
 \For{$i \gets 0$ to $n$}{
 $attackvector \gets$ []\\
 \For{$j \in |X.cols|$}{
    $r \gets \mathcal{N}(0, 3)$\\
    $attackvector.add(\mu[j] + r \times \sigma[j])$
 }
 $X'.add(attackvector)$
 }
 \caption{$\texttt{population\_attack}(X, n)$}
 \label{attack}
\end{algorithm}

\textbf{The adversarial environments.} We implemented two adversarial environments. First, the zero-effort adversarial environments, one of the most studied adversarial environments in the literature. Likely because it is the easiest to implement. The feature vectors of the users other than genuine users serve as the attack vectors in this environment. The second commonly studied adversarial environment is the population-based attack in which the impostor samples are derived from all the users in a given database. In the phone case, we were able to find a separate public dataset (Serwadda-Touch). Thus, we used BBMAS-Touch as well as the Serwadda-Touch dataset for population attack. In contrast, we failed to find such a dataset for the tablet environment. Consequently, we used the same BBMAS-Touch for the population attack. The process of computing the population feature vectors from a dataset is described through Algorithm \ref{attack}. As shown, we generated $n$ population attack vectors, using the formula = $\mu + r \times \sigma$ where $r = \mathcal{N}(0, 3)$. 
\vspace{0.1in}

\textbf{Training V-TCAS.} We followed the strategy widely implemented in the literature to train the Vanilla authentication framework. To train the selected classifiers for user $u_i$, we labeled the feature vectors belonging to $u_i$ as genuine and the feature vectors belonging to the other users than $u_i$ as impostors. We conducted five-fold cross-validation to find the hyperparameters that resulted in the highest balanced accuracy. The models were retrained using the best hyperparameters and pickled for testing. 
\vspace{0.1in}

\textbf{Training G-TCAS.} 
GANs have been extensively used to generate synthetic data points \cite{goodfellow2014generative}. It consists of Generator ($G$) and Discriminator ($D$). $G$ captures the data distribution, whereas $D$ estimates the probability of generated data belonging to $G$. $G$ is trained until $D$ exhibits an acceptable level of error. The idea behind GANs is to consider $G$ and $D$ as two players of a min-max game with value function $V (D, G)$ represented in Equation \ref{eq:eq9}.
\begin{equation}
    E_1 = \mathbb{E}_{\boldsymbol{x} \sim p_{\text {data }}(\boldsymbol{x})}[\log D(\boldsymbol{x})]
\end{equation}
\begin{equation}
    E_2 = \mathbb{E}_{\boldsymbol{z} \sim p_{\boldsymbol{z}}(\boldsymbol{z})}[\log (1-D(G(\boldsymbol{z})))]
\end{equation}
\begin{equation}
\label{eq:eq9}
\min_{G} \max_{D} V(D, G)= E_1 + E_2
\end{equation}
To learn the distribution $p_g$ over data $x$, a prior on input noise variables $p_z(z)$ is defined and $G(z; \theta_g)$  represent a mapping to data space. Where $G$ is a function with parameters $\theta_g$.
Second layer $D(x; \theta_d)$ outputs a single scalar. Here, $D(x)$ represents the probability that $x$ came from the data and not from $p_g$. We train $D$ to maximize the probability of assigning the
correct label to both training examples and samples from $G$. We simultaneously train $G$ to minimize
$\log (1 - D(G(z)))$.

The combination of GANs generated data and the actual dataset can make the authentication models more resilient to attacks. This hypothesis motivated us to implement a GAN-assisted TCAS. Figure \ref{adversarialtouch} demonstrates the block diagram of the framework. To summarize, we used two GANs to generate both legitimate and adversarial feature vectors. We customized the implementation of GAN provided at \cite{GANimplementation} to fit our requirements. In particular, we tuned the parameters (epoch, bath\_size, and learning\_rate) to obtain the accuracy beyond $93\%$ for both GANs. We refer to these GANs as Legitimate-GAN and Adversarial-GAN, respectively. We generated synthetic samples for each user separately. The Legitimate-GAN input was the genuine feature vectors, while the input for the Adversarial-GAN was the impostor feature vectors (\ie feature vectors from users other than the genuine user).  

The GANs did not perform well when the number of requested synthetic feature vectors was too low or too high. To find the optimal number of feature vectors generated, we randomly tested many values between $[50, 1000]$ and found that $n=250$ was optimal.  

\begin{table*}[htp!]
\small
\centering
\caption{Authentication results for both Vanilla TCAS (V-TCAS) and GAN TCAS (G-TCAS) are presented in terms of mean FAR, FRR, and HTER under zero-effort and population-based adversarial environments, for both smartphone and tablet devices. The performance of both V-TCAS and G-TCAS are comparable for the zero-effort adversarial environment in almost all the settings. However, the performance of the V-TCAS degraded significantly under the population-based adversarial environment while G-TCAS showed more resilience to the population-attack.}
\begin{tabular}{|c|c|c|c|c|c|c|c|c|c|c|}
\hline
\multicolumn{2}{|c|}{\multirow{2}{*}{\textbf{Device}}}                                                                                 & \multirow{2}{*}{\textbf{Metric}} & \multicolumn{2}{c|}{\textbf{SVM}}    & \multicolumn{2}{c|}{\textbf{RForest}} & \multicolumn{2}{c|}{\textbf{MLP}}    & \multicolumn{2}{c|}{\textbf{XGBoost}} \\ \cline{4-11} 
\multicolumn{2}{|c|}{}                                                                                                                 &                                  & \textbf{V-TCAS} & \textbf{G-TCAS} & \textbf{V-TCAS}  & \textbf{G-TCAS} & \textbf{V-TCAS} & \textbf{G-TCAS} & \textbf{V-TCAS}  & \textbf{G-TCAS} \\ \hline
\multirow{7}{*}{\textbf{Phone}}  & \multirow{3}{*}{\textbf{Zero-effort}}                                                               & \textbf{FRR}                     & $0.03$             & $0.03$              & $0.03$              & $0.03$              & $0.04$             & $0.04$              & $0.03$              & $0.03$              \\ \cline{3-11} 
                                 &                                                                                                     & \textbf{FAR}                     & $0.09$             & $0.12$              & $0.08$              & $0.07$              & $0.08$             & $0.08$              & $0.07$              & $0.07$              \\ \cline{3-11} 
                                 &                                                                                                     & \textbf{HTER}                    & $0.06$             & $0.07$              & \textbf{0.05}              & \textbf{0.05}              & $0.06$             & $0.06$              & \textbf{0.05}              & \textbf{0.05}            \\ \cline{2-11} 
                                 & \multirow{2}{*}{\textbf{\begin{tabular}[c]{@{}c@{}}Population \\ (Same)\end{tabular}}}      & \textbf{FAR}                     & \textbf{0.23}             & \textbf{0.11}              & $0.29$              & \textbf{0.10}              & \textbf{0.28}            & $0.24$              & $0.24$              & $0.19$              \\ \cline{3-11} 
                                 &                                                                                                     & \textbf{HTER}                    & $0.11$             & $0.05$              & $0.14$              & $0.05$              & $0.14$             & $0.12$              & $0.12$              & $0.09$              \\ \cline{2-11} 
                                 & \multirow{2}{*}{\textbf{\begin{tabular}[c]{@{}c@{}}Population \\ (Different)\end{tabular}}} & \textbf{FAR}                     & $0.29$             & \textbf{0.19}              & \textbf{0.26}             & \textbf{0.13}              & $0.36$             & $0.28$              & \textbf{0.28}             & $0.27$              \\ \cline{3-11} 
                                 &                                                                                                     & \textbf{HTER}                    & $0.14$             & $0.09$              & $0.13$              & $0.06$              & $0.18$             & $0.14$              & $0.14$              & $0.13$              \\ \hline
\multirow{5}{*}{\textbf{Tablet}} & \multirow{3}{*}{\textbf{Zero-effort}}                                                               & \textbf{FRR}                     & $0.05$             & $0.02$              & $0.06$              & $0.03$             & $0.05$             & $0.04$              & $0.05$              & $0.03$              \\ \cline{3-11} 
                                 &                                                                                                     & \textbf{FAR}                     & $0.07$             & $0.12$              & $0.07$              & $0.08$              & $0.09$             & $0.09$              & $0.07$              & $0.07$              \\ \cline{3-11} 
                                 &                                                                                                     & \textbf{HTER}                    & \textbf{0.06}            & $0.07$              & \textbf{0.06}              & \textbf{0.05}             & $0.07$             & $0.06$              & \textbf{0.06}              & \textbf{0.05}              \\ \cline{2-11} 
                                 & \multirow{2}{*}{\textbf{\begin{tabular}[c]{@{}c@{}}Population\\ (Same)\end{tabular}}}       & \textbf{FAR}                     & \textbf{0.19 }            & \textbf{0.12}              & $0.38$              & $0.13$              & $0.36$             & $0.18$              & $0.4$               & $0.19$              \\ \cline{3-11} 
                                 &                                                                                                     & \textbf{HTER}                    & $0.09$             & $0.06$              & $0.19$              & $0.06$              & $0.18$             & $0.09$              & $0.20$               & $0.09$              \\ \hline
\end{tabular}
\label{tabresults1}
\end{table*}
\vspace{0.1in}

\textbf{Performance evaluation.} The performance of the implemented TCAS was evaluated using three measures, False Accept Rate (FAR--the percentage of successful impostor attempts), False Reject Rates (FRR--the percentage of failed genuine attempts), and Half Total Error Rate (HTER--an average of FAR and FRR). Another measure that has been widely used to report the test performance of TCAS is Equal Error Rate (EER). However, researchers have advised using HTER instead of EER to report the \textit{test} performance of a biometric system primarily because EER is computed by varying the threshold. The threshold, however, cannot and should not be adjusted during testing \cite{bengio2002confidence}. The attack's impact can be measured by the increase in the False Accept Rates (FAR).  

\begin{figure*}[htp]
\centering
\begin{tabular}{cccc}
\subfigure{\epsfig{file=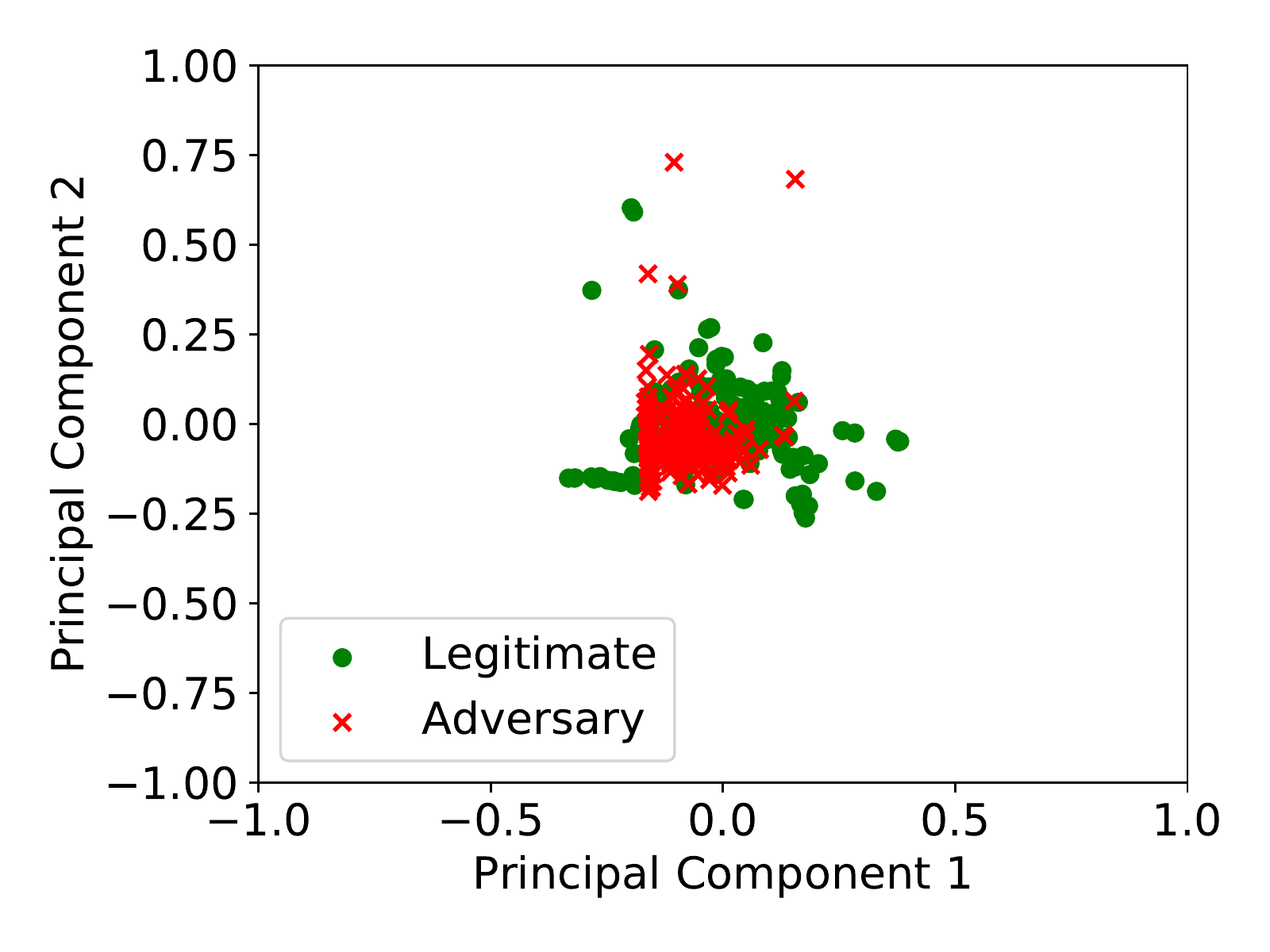, width=1.65in, height=1.25in} \label{nonGAN3}}
\subfigure{\epsfig{file=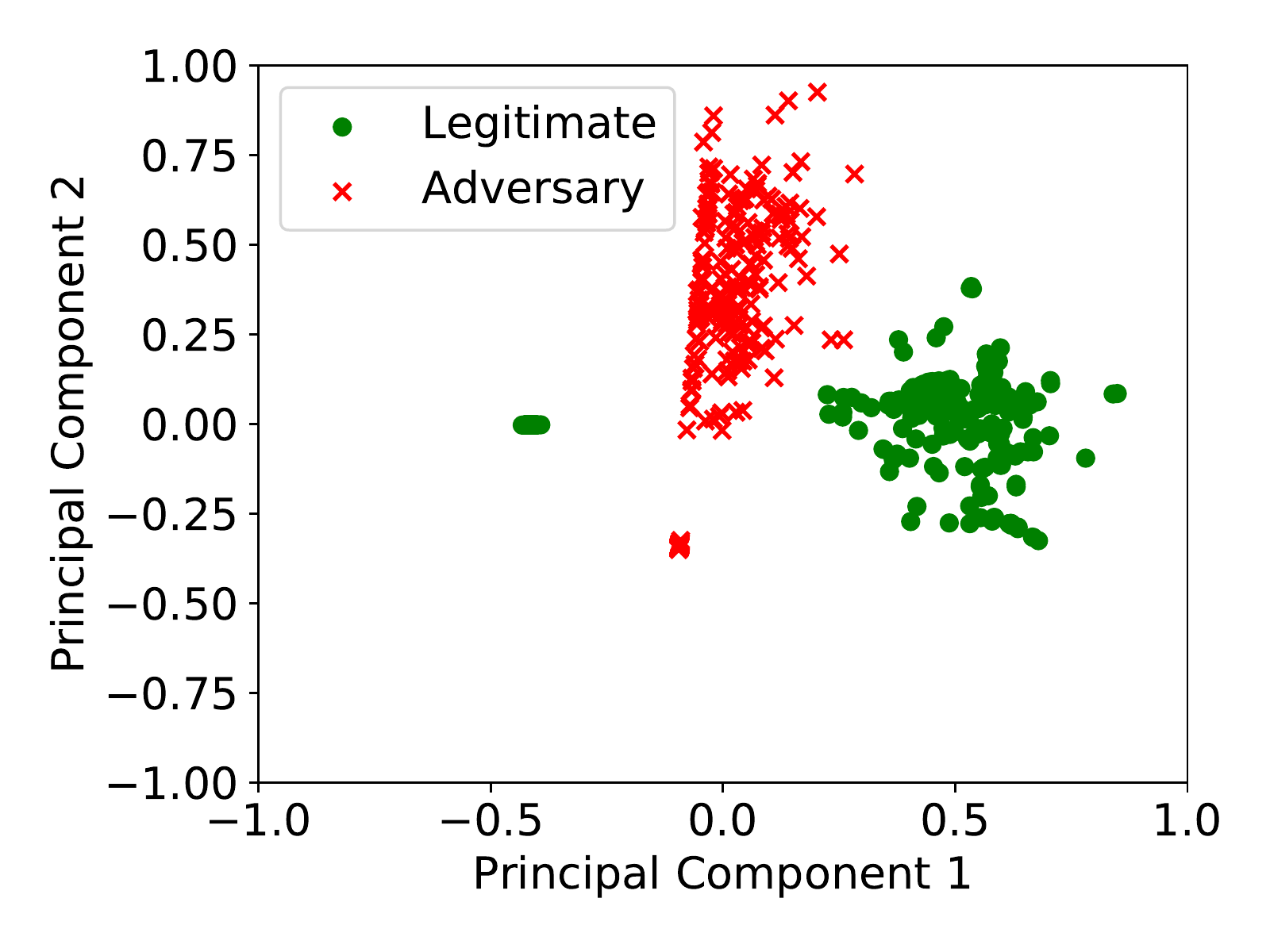, width=1.65in, height=1.25in}
\label{withGAN3}}
\subfigure{\epsfig{file=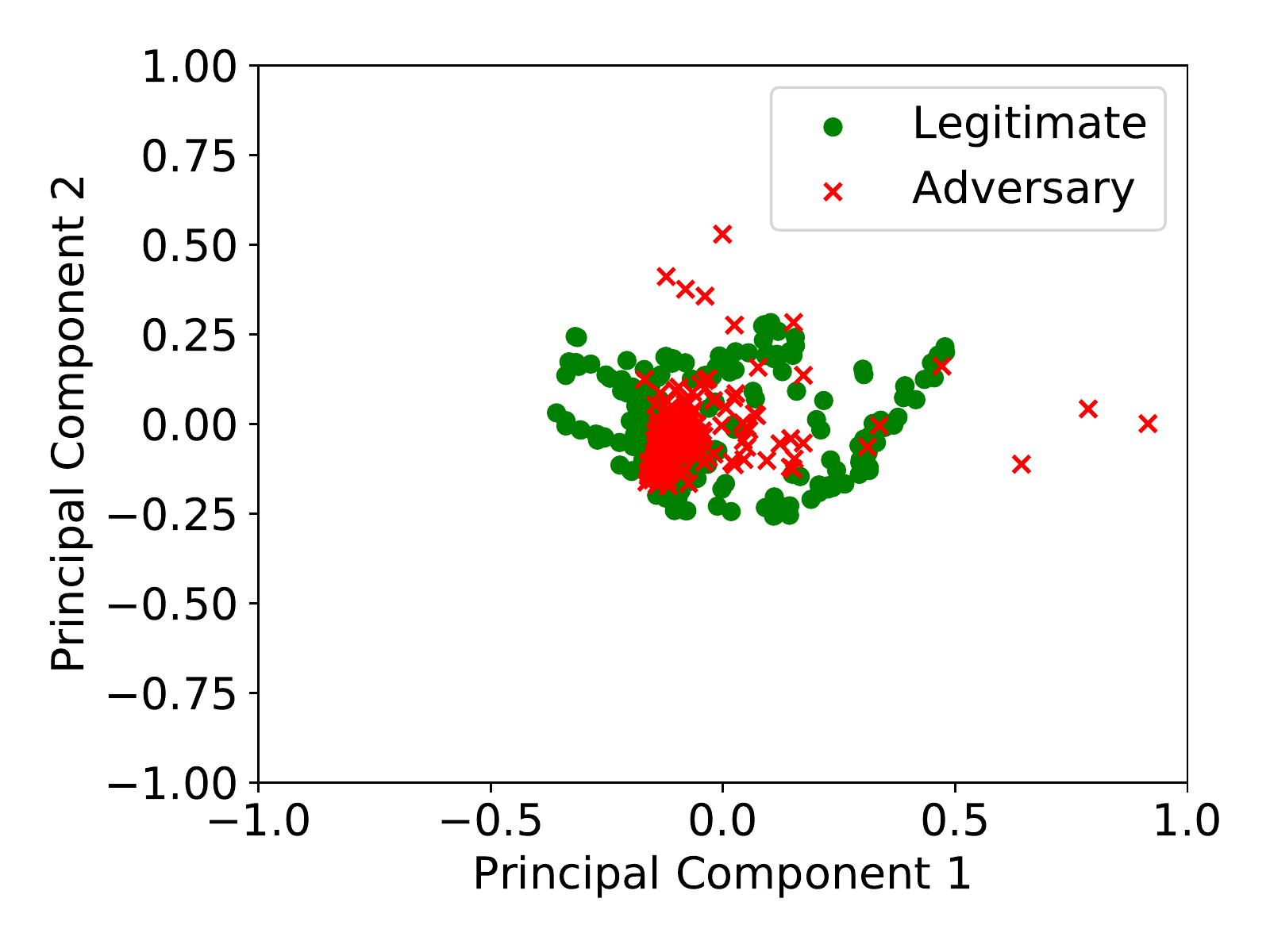, width=1.65in, height=1.25in} 
\label{nonGAN4}}
\subfigure{\epsfig{file=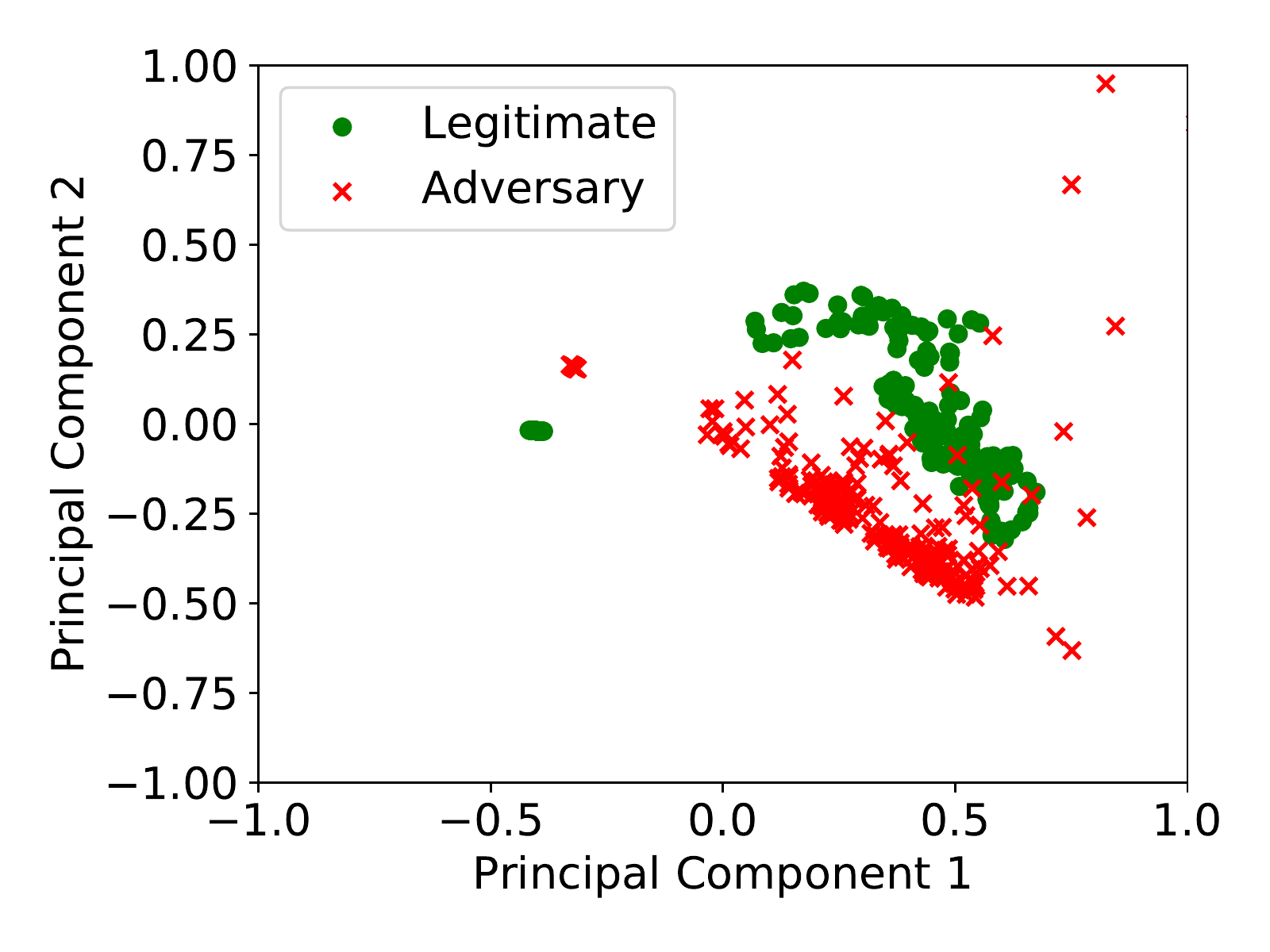, width=1.65in, height=1.25in}
\label{withGAN4}}
\end{tabular}
\caption{Demonstrating the feature space separation between V-TCAS vs G-TCAS using top two components that were obtained by conducting the Principal Component Analysis (PCA) for two random users.} 
\label{figgan}
\end{figure*}

\section{Results and Discussion} 
\label{ResultsAndDiscussion}
\subsection{Performance Analysis of V-TCAS} Table \ref{tabresults1} presents the performance of Vanilla TCAS (V-TCAS) in terms of FAR, FRR, and HTER under zero-effort for both smartphone and tablet devices. V-TCAS's performance under the population-based adversarial environment is reported using only FAR, and HTER; the FRRs remain the same as of the zero-effort adversarial environment. 
\vspace{0.1in}

\textbf{Smartphone.} Both Random Forest and XGBoost achieved $5\%$ HTERs which was better than SVM and MLP ($6\%$ HTERs) under the zero-effort adversarial setup. On the other hand, SVM ($14\%$ increase in the FAR) and XGBoost ($17\%$ increase in FAR) showed more resilience to the population-based adversarial setup created using the same dataset than Random Forest ($21\%$ increase in FAR) and MLP ($20\%$ increase in FAR). 

\textbf{Tablet.} The performance of Random Forest and XGBoost rounded to $5\%$ HTER, which was better than MLP ($6\%$ HTER) and SVM ($7\%$ HTER). SVM was the most resilient ($12\%$ increase in the FAR) classifier compared to MLP ($27\%$), Random Forest ($31\%$), and XGBoost ($33\%$). 

We can conclude that SVM was the most resilient classifier, irrespective of the device. 

\subsection{Performance Analysis of G-TCAS}
Table \ref{tabresults1} presents the G-TCAS's performance in terms of FAR, FRR, and HTER under the zero-effort for both smartphone and tablet devices. G-TCAS's performance under the population-based adversarial environment is reported using only FAR and HTER; the FRRs remain the same as of the zero-effort adversarial environment. 
\vspace{0.1in}

\textbf{Smartphone.} GTCAS achieved similar performance to V-TCAS for the majority of the classifiers. Random Forest ($5\%$ HTER) and XGBoost ($5\%$ HTER) did better than SVM ($7\%$ HTER) and MLP ($6\%$ HTER) under the zero-effort adversarial setup. However, SVM ($-1\%$ increase in FAR) and Random Forest ($3\%$ increase in FAR) showed more resilience to the population-based adversarial setup created using the same dataset than MLP ($16\%$ increase in FAR) and XGBoost ($12\%$ increase in FAR). The same phenomenon holds for the population-based adversarial setup created using a different dataset as RForest ($6\%$) and SVM ($7\%$), which showed more resilience than XGBoost ($20\%$) and MLP ($28\%$). 
\vspace{0.1in}

\textbf{Tablet.} The performance of SVM and Random Forest rounded to $5\%$ HTER, which was better than MLP ($7\%$ HTER) and XGBoost ($7\%$ HTER). SVM ($0\%$ increase in FAR) and Random Forest ($5\%$ increase in FAR) were most resilient compared to MLP ($9\%$) and XGBoost ($12\%$).

\subsection{V-TCAS vs. G-TCAS} Overall, the results table suggest that G-TCAS performed much better under a stringent (population-based) adversarial environment than V-TCAS. GAN-generated synthetic samples (both legitimate and adversarial) included in the authentication pipeline helped draw a better boundary. To analyze the idea, we plotted the top two principal components for a random user (a sample is shown in Figure \ref{figgan}). We observed that the G-TCAS helped separate the classes better, making the implementation more robust than V-TCAS. 

\section{Conclusion and Future Work}
\label{secConclusionAndFutureWork}
We proposed a novel framework to implement TCAS, which showed more resilience to the population-based adversarial attacks than the widely studied TCAS designs. SVM-based implementations consistently performed better than the other classifiers both in terms of accuracy and resilience.

This work is limited to touch-biometrics, ML models, and two datasets; we would like to expand the study across different biometrics (\eg, gait, keystroke), learning frameworks (\eg, DL models, and one-class paradigm), and a variety of datasets. Additionally, we aim to explore whether the proposed defense mechanism works under more rigorous attack environments \eg, imitation-based attacks. 

\section{Acknowledgment} We thank the anonymous reviewers for their insightful
feedback. Rajiv Ratn Shah was partly supported by the Infosys Center for Artificial Intelligence and the Center of Design and New Media at IIIT Delhi, India.

{\small
\bibliography{camera_ready}
}

\end{document}